\ifx\pdfoutput\undefined
	\documentclass[aps,prd,a4paper,twoside,onecolumn,english,superscriptaddress,
	noshowkeys,nofootinbib,amssymb,amsmath,amsfonts,dvips]{revtex4}
\else
	\documentclass[aps,prd,a4paper,twoside,onecolumn,english,superscriptaddress,
	noshowkeys,nofootinbib,amssymb,amsmath,amsfonts,pdftex]{revtex4}
\fi
\usepackage{babel}
\usepackage{graphicx}

\begin{document}

\pagestyle{empty}
\setcounter{page}{263}

\title{On the possible role of massive neutrinos in cosmological structure formation}

\author{\firstname{Massimiliano} \surname{Lattanzi}}
\affiliation{ICRA --- International Center for Relativistic Astrophysics.}
\affiliation{Dipartimento di Fisica, Universit\`a di Roma ``La 
Sapienza'', Piazzale Aldo Moro 5, I-00185 Roma, Italy.}

\author{\firstname{Remo} \surname{Ruffini}}
\affiliation{ICRA --- International Center for Relativistic Astrophysics.}
\affiliation{Dipartimento di Fisica, Universit\`a di Roma ``La 
Sapienza'', Piazzale Aldo Moro 5, I-00185 Roma, Italy.}

\author{\firstname{Gregory} \surname{Vereshchagin}}
\affiliation{ICRA --- International Center for Relativistic Astrophysics.}
\affiliation{Belorussian State University, Theoretical Physics Department, Skorina ave.  4, 220050, Minsk, Republic of Belarus.}

\begin{abstract}
In addition to the problem of galaxy formation, one of the greatest open questions of cosmology is represented by the existence of an asymmetry between matter and antimatter in the baryonic component of the Universe. We believe that a net lepton number for the three neutrino species can be used to understand this asymmetry. This also implies an asymmetry in the matter-antimatter component of the leptons. The existence of a nonnull lepton number for the neutrinos can easily explain a cosmological abundance of neutrinos consistent with the one needed to explain both the rotation curves of galaxies and the flatness of the Universe. Some propedeutic results are presented in order to attack this problem.
\end{abstract}


\maketitle
\tableofcontents
\section{Evidence for dark matter and the possible role of neutrinos}

The most popular model of the Universe being currently discussed in the literature is one usually indicated by the ``$\Lambda$CDM model'' (see e.g. E.W. Kolb, in these same proceedings \cite{kolbBr}), which implies that almost 75\% of the energy density of the Universe is due to a cosmological term, while only 1\% is due to  neutrinos \cite{MAP1}--\cite{MAP13}. This result seems to be in conflict with our current knowledge of the rest of physics. We will give in the following some propedeutical considerations to reconsider this problem.

Many theoretical considerations and observational facts make it clear (see \cite{kolb} and references therein) that luminous matter alone cannot account for the whole matter content of the Universe. Among them there are the considerations on cosmological nucleosynthesis \cite{1991R}-\cite{Bo85} as well as the measurements of the the cosmic background radiation (CBR) anisotropy spectrum \cite{BOOM1}-\cite{BOOM2}. In both cases the fit is consistent with a cosmological model in which just a fraction smaller than 10\% of the total density is due to baryons. 

Strong evidence for the presence of dark matter is directly given by the rotation curves of galaxies \cite{Sa87}. If we assume for simplicity a spherical or ellipsoidal mass distribution for a galaxy, the orbital velocity at a radius $r$ is given by Newton's equation of motion:
\begin{equation}
\label{Newt}
v^2=\frac{GM(r)}{r}
,
\end{equation}
where $G$ is the gravitational constant and $M(r)$ is the mass contained in a sphere of radius $r$. The peculiar velocity of stars beyond the visible edge of the galaxy should then decrease as $1/r$. What is observed instead is that the velocity stays nearly constant with increasing $r$. This requires a halo of invisible dark matter to be present outside the radius of the visible matter. From observations it follows that the halo radius can be 10 times larger than the radius of visible part of the galaxy. Then from Eq.~(\ref{Newt}) it follows that $M_{halo}$ is at least 10 times larger than the galactic mass $M_{gal}$.
We now assume that galactic halos are composed of neutral fermions of mass $m_x$ and apply a gravitational Thomas-Fermi model to this system. The equation for the dimensionless gravitational potential $\chi$ as a function of a radial coordinate $x$ is
\begin{equation}
\label{LEeqn}
\frac{d^2\chi}{dx^2}=-\frac{\chi^{3/2}}{\sqrt{x}}
,
\end{equation}
with the initial condition $\chi(0)=0$. For any number of particles $N$ this equation can be used to compute the radius $R$ of the system. Since the total mass is $M=Nm_x$, this defines a relation between $M$, $R$ and $m_x$ which  allows one to estimate $m_x$, since both the value of the total mass and radius are known. Using for the radius of the halo $R_{halo}=10R_{gal}$ and for the mass of the halo $M_{halo}=10M_{gal}$ we obtain the particle mass $m_x\approx4$eV. This value, although obtained by a simple argument, is in good agreement with the recent constraints on the electron neutrino mass obtained from the spectrum of tritium beta decay ($m_\nu<2.5$ eV \cite{mLab}).

This agreement can even be improved if one takes into account different families of neutrinos. In this case, the effect of the additional degrees of freedom can be expressed in terms of the effective mass $m_{eff}=\left(\sum_{i=1}^n m_i^4\right)^{1/4}$. If we consider three families with nearly the same mass, they provide a factor $3^{1/4}\simeq1.32$ to improve the bound to $m_\nu\simeq3\,$eV.

Neutrinos were considered as the best candidate for dark matter about thirty years ago. Indeed, it was shown by Gerstein and Zel'dovich in 1966 \cite{GZ} that if these particles have a small mass $m_\nu\sim30\,$eV, they provide a large energy density contribution up to the value $\Omega_\nu\sim1$. It is in effect easy to show that the density parameter of a single family of mass $m_\nu\ll 1$~Mev would be, assuming a null chemical potential:
\begin{equation}
\Omega_\nu h^2=\frac{m_\nu}{93\,\mathrm{eV}}
,
\label{GZ}
\end{equation}
where $h$ is the Hubble constant in units of 100~km~sec$^{-1}$~Mpc$^{-1}$. The generalization to the case of a nonnull chemical potential is (see e.g. Ruffini \& Song \cite{1986R}):
\begin{equation}
\Omega_\nu h^2=\frac{m_\nu}{93\,\mathrm{eV}}A(\xi)
,
\label{GZ2}
\end{equation}
where $\xi$ is the dimensionless chemical potential (so called degeneracy parameter) at decoupling and
\begin{equation}
A(\xi)\equiv
\frac{1}{4\eta(3)}\left[\frac13\,|\xi|^3+4\eta(2)|\xi|
+4\sum_{k=1}^\infty(-1)^{k+1}\,\displaystyle\frac{e^{-k|\xi|}}{k^3}\right]
.
\label{Axi}
\end{equation}
Here $\eta(n)$ denotes the Riemann $\eta$ function of the index $n$. This allows to obtain even higher values of $\Omega_\nu$. In fact we have carried out a more detailed analysis, taking into account three different neutrino flavours with the same mass but different chemical potentials. Using the limits that primordial nucleosynthesis imposes on the degeneracy parameters of the $e$, $\mu$ and $\tau$ neutrinos \cite{1991R}, we found that if $m_\nu\simeq2$~eV and $\xi_e\simeq0.4$ then $\Omega_\nu\simeq1$ \cite{Latt}.

However, in 1979 Tremaine and Gunn \cite{TrGu} claimed that massive neutrinos cannot be considered as a dark matter candidate. Their paper was very influential and turned most cosmologists away from neutrinos as cosmologically important particles. In their paper, Tremaine and Gunn establish an upper limit to the neutrino mass of $m_\nu\lesssim 1.2$~eV. They obtain this limit from arguments based on the velocity dispersion within galaxies and clusters, under the hypothesis that $m_\nu\lesssim 1$~MeV, so that neutrinos are ultrarelativistic at decoupling. They also obtain a lower bound $m_\nu\gtrsim20$~eV from an argument based on phase space density considerations and on the rotation curves of galaxies. In this very strange situation with wildly contradictory constraints, they possibly see a way out by avoiding the fact that neutrinos are ultrarelativistic at decoupling and they conclude that massive galactic halos cannot be composed of stable neutral leptons of mass $\lesssim1$~MeV. 

While the Gunn and Tremaine result deeply influenced astrophysicists against the possible role neutrinos in cosmology, especially in the U.S.A., in 1977 Lee and Weinberg \cite{LeeWe} turned their attention to massive neutrinos with $m_\nu>2$~GeV, showing that such particles could provide a large contribution to the energy density of the Universe, in spite of a much smaller value of their number density. This paper was among the first to consider very massive particles as candidates for dark matter. This very interesting work, together with the Gunn and Tremaine purported difficulties for the neutrino scenario, induced some cosmologists to turn their attention to very massive particles, thus marking the birth of cold dark matter models. 

A clear counterexample to the Gunn and Tremaine bound, which was indeed derived from a nontransparent mixture of quantum limits on classical Maxwell-Boltzmann statistics, was given by Gao and Ruffini \cite{1980GR}. They established a different upper bound on the neutrino mass from the assumption that galactic halos are composed of degenerate neutrinos: $m_\nu\lesssim15$ eV. This result was further developed and confirmed by Arbolino and Ruffini \cite{1988AR}. They explicitly showed that rotation curves for galaxies in perfect agreement with the observations can be obtained for neutrino masses of the order of 9~eV. This limit could be lowered further if semidegenerate configurations for the neutrino halo were to be considered (see e.g. Merafina and Ruffini \cite{MR89}-\cite{MR90}); Ingrosso, Merafina, Ruffini and Strafella \cite{IMR92}).

Today, quite apart from the rotation curves of galaxies, the recent determination of the neutrino masses would appear to be in contradiction with an assumption of $\Omega_\nu\sim1$. However, this is only an apparent difficulty, since for semidegenerate distributions (see Eq.~(\ref{GZ2})) this equality can indeed be fulfilled and important consequences on the matter-antimatter asymmetry in the leptonic component of the Universe  can be inferred.

One of the most interesting features of neutrino cosmology is that they establish a natural cutoff for the largest possible structure in the Universe, related to the maximum value of the Jeans mass when the neutrinos become nonrelativistic:
\begin{equation}
\label{Jm1}
M_J(z_{nr})=1.475\cdot\;
10^{17}M_{\odot}g_{\nu}^{-\frac{1}{2}}N_{\nu}^{-\frac{1}{2}}
\left(\frac{m_{\nu}}{10eV}\right)^{-2}
A(\xi)^{\frac{5}{4}}B(\xi)^{\frac{3}{4}}
,
\end{equation}
where $z_nr$ is the redshift at which neutrinos enter the nonrelativistic regime, $g_\nu$ is the number of quantum degrees of freedom, $N_\nu$ is the number of neutrino families, $A(\xi)$ is as defined in Eq.~(\ref{Axi}), and 
\begin{equation}
B(\xi)\equiv
\frac{1}{48\eta(5)}\left[\frac{1}{5}\xi^5+8\eta(2)\xi^3 
+48\eta(4)\xi +48\sum_{n=1}^{\infty}(-1)^{n+1} \frac{e^{-n\xi}}{n^5}\right]
.
\end{equation}
This mass appears to be essential in determining the upper cutoff for a possible fractal structure of the Universe. Real difficulties still exist today in understanding the details of the fragmentation of these masses of $10^{17}M_\odot$ and the development of smaller structures all the way down to galaxies. In this lecture we illustrate some of this basic problems which still need additional work before a detailed correspondence with observations can be obtained. 
\section{Large scale structure}

\subsection{The cosmological principle}

There have been three distinct moments in the development of the so called \emph{cosmological principle} which is at the very basis of our approach to the analysis of the Universe. The first formulation of the cosmological principle can be simply stated:
\begin{center}
\begin{equation}
\mathrm{\emph{All the events in the Universe are equivalent.}}
\end{equation}
\end{center}
Such a cosmological principle was enunciated a few years after the introduction of the field equations of general relativity by Einstein himself \cite{Ei17} in the quest for visualizing a Universe the most democratic with respect to any special point and any possible moment of time: a Universe everlasting in time and totally homogenous in the spatial directions. No solution fulfilling such a cosmological principle could be found, and Einstein was so strongly confident of the validity of this principle that he modified his field equations of general relativity by introducing a cosmological constant $\Lambda$. George Gamow said that Einstein later on considered that the biggest mistake in his life.
 
It was through the work of Alexander Alexandrovich Friedmann \cite{Fr22}-\cite{Fr24} that a new cosmological principle was advanced:
\begin{center}
\begin{equation}
\mathrm{\emph{All the points in the Universe are equivalent.}}
\end{equation}
\end{center}
As long as we look at our `neighbour' Universe, this statement is certainly false, because the distribution of matter is far from homogeneous: there are planets, stars, and going to larger scales, galaxies and clusters of galaxies separated by almost empty regions.
However, the Friedmann principle should apply when we average this distribution over a volume containing a large enough number of galaxies. For such a spatially homogeneous Universe Friedmann \cite{Fr22} found in 1922 explicit analytic solutions of the Einstein equations of general relativity. A remarkable property of these solutions is that they describe a non-static Universe. At that time, there was no observational evidence for the temporal evolution of the whole Universe. The first evidence came in 1929 from the observation by Hubble \cite{Hu29} of the recession of the nebulae. Hubble was the first trying to study the spatial distribution of objects as large as the galaxies, at that time thought to be the largest self-gravitating systems to exist. The Hubble law, interpreted within the framework of Friedmann cosmology, implied that the galaxy distribution is close to homogenous on the large-scale average \cite{We52}--\cite{Le31b}. 

It was through the above mentioned work of Hubble together with the later remarkable work of George Gamow and his collaborators (1946--1949) \cite{Ga46}-\cite{AH49} who postulated an initially hot Universe, and the detailed work of Fermi and Turkievich in the same years \cite{AH50} introducing the first computation of cosmological nucleosynthesis, that the Friedmann Universe has grown to become the standard paradigm in cosmology following the discovery of the CBR by Penzias \& Wilson in 1965 \cite{Pe65}.\\
\begin{figure}
\begin{center}
\includegraphics[width=0.8\hsize,clip]{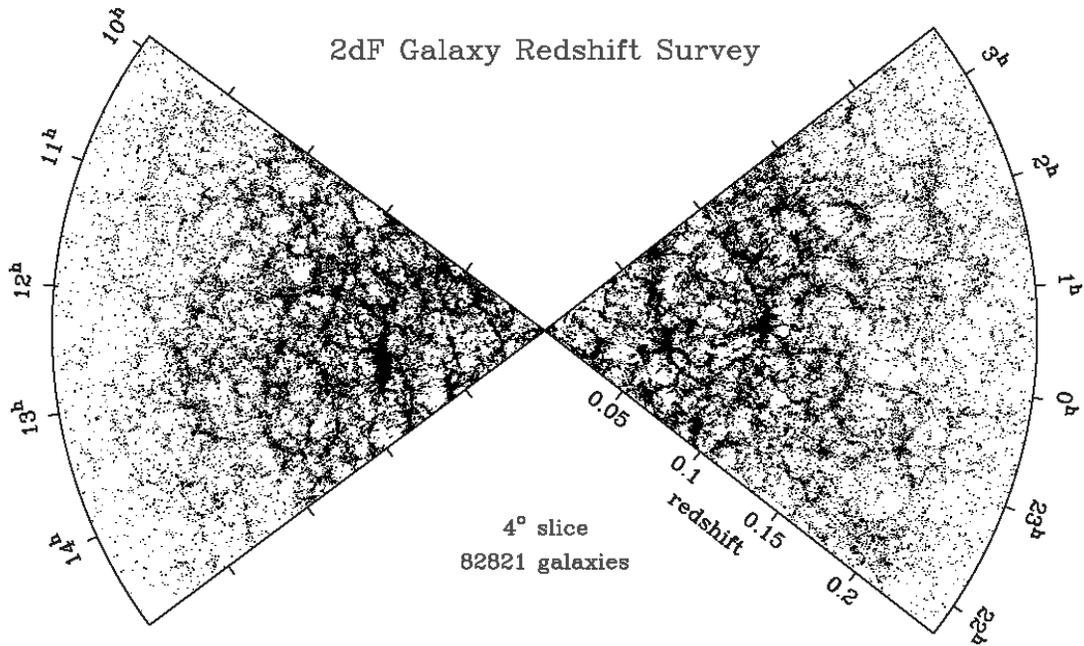}
\caption{The distribution of galaxies in the 2dFGRS (from \cite{2dF}). Courtesy of J.A. Peacock and the 2dFGRS Team.}
\label{fig:2dF}
\end{center}
\end{figure} 
In effect, one of the strongest predictions of Big Bang model is the presence of a background microwave radiation, a relic of the early Universe. This radiation is highly isotropic, reflecting through the coupling with matter the high isotropy and homogeneity of the primeval plasma. This tells us that the cosmological principle, and then the Friedmann picture, safely apply to the early Universe. Homogeneity on very large scales is confirmed by present day observations of in particular:
\begin{itemize}
\item X-ray background \cite{Xray},\\
\item radio sources \cite{Radio},\\
\item gamma ray bursts distribution \cite{batse4b},\\
\item galaxies and clusters of galaxies \cite{2dF}.
\end{itemize}

So much so for the very large scales, but what about structures like galaxies and clusters of galaxies? More and more they appear to be distributed without any apparent homogeneity, but on the contrary showing regularities in an apparent hierarchical distribution of galaxies, clusters of galaxies and superclusters of galaxies separated by large voids \cite{CP}--\cite{gaite} (see Fig.~\ref{fig:2dF}). Slowly but more and more clearly the presence of a fractal distribution in the Universe has started to surface and with it a new cosmological principle which can be simply expressed:
\begin{center}
\begin{equation}
\emph{All the observers in the Universe are equivalent.}
\label{CP3}
\end{equation}
\end{center}
We shall recall in the following a few basic points which are essential in arriving at this new principle and make possible the verification of its possible validity.

\subsection{Two-point correlation function}

The statistical description of clustering is based on the concept of correlation, namely, more precisely, on the probability of finding an object in the vicinity of another one. The standard way to quantify this probability is to define the two-point correlation function $\xi(\vec x)$ \cite{peebles}.

Consider a distribution of objects in space described by the number density function $n(\vec x)$.
The probability that an object is found in an infinitesimal volume $\delta V$ centered around the point $\vec x$ is proportional to the volume itself:
\begin{equation}
\delta P\propto \delta V
.
\end{equation}
In the absence of structure, the joint probability of finding two objects in two different infinitesimal volumes $\delta V_1$ and $\delta V_2$, centered respectively around $\vec x_1$ and $\vec x_2$, is given by the product of the two probabilities:
\begin{equation}
\delta P=\delta P_1\delta P_2\propto\delta V_1\delta V_2
.
\end{equation}
On the other hand, if objects have a tendence to cluster, we will find an excess probability:
\begin{equation}
\label{xi1}
\delta P\propto\delta V_1\delta V_2\cdot(1+\xi(\vec x_1, \vec x_2))
.
\end{equation}
According to the cosmological principle, we don't expect the correlation function to depend either on the position or on the direction, but only on separation beetween volumes: $\xi(\vec x_1, \vec x_2)=\xi(r_{12})$, where $r_{12}\equiv |\vec x_1-\vec x_2|$.

An equivalent definition of the two-point correlation function is the following:
\begin{equation}
\label{xi2}
\xi(r_{12})=<\delta(\vec x_1)\delta(\vec x_2)>
,
\end{equation}
where $<...>$ denotes averaging over all pairs of points in space separated by a distance $r_{12}$, and $\delta(\vec x)\equiv(n(\vec x)-\bar n)/\bar n$. 

\subsection{Observed galaxy distribution}

Observational data coming from galactic surveys are usually expressed in the form of a correlation function  $\xi(\pi,\sigma)$ in redshift space, where $\pi$ is a separation along the line of sight and $\sigma$ is a angular separation on the plane of the sky between two galaxies. It is then possible to obtain the real-space correlation function $\xi(r)$; this step is never a trivial one, but we will not go into details since it is beyond the scope of this review. 

Peebles \cite{peebles} has shown that the distribution of galaxies can be described by a two point correlation function with a simple power law form:
\begin{equation}
\xi_g(r)=\left(\frac{r}{r_g}\right)^{-1.77}, \;\;\;\; 
r<10h^{-1}\mathrm{Mpc}
,
\end{equation}
where $h$ is the present day Hubble parameter measured in units of $100\,\frac{km}{s\,Mpc}$.
The correlation length $r_g$ determines the typical distance between objects. For galaxies, it has been estimated to be $\simeq 5h^{-1}\mathrm{Mpc}$.

For clusters of galaxies the same power law was  found first by Bahcall and Soneira \cite{BahSon} and then Klypin and Kopylov \cite{KlyKop}
\begin{equation}
\xi_c(r)
=\left(\frac{r}{r_c}\right)^{-1.8}, \;\;\;\; 
  5h^{-1}<r<150h^{-1}\mathrm{Mpc}
\end{equation}
with different correlation lengths, namely $r_c\simeq25h^{-1}\mathrm{Mpc}$.
Furthermore, Bahcall and Burgett \cite{BahBur} found a correlation function for superclusters of galaxies with the same power law.

Recent observations support these conclusions.
The first results from the Sloan Digital Sky Survey (SDSS) on galaxy clustering \cite{Zehavi} for about 30,000 galaxies give a real-space correlation function of
\begin{equation}
\xi_g(r)
=\left(\frac{r}{r_0}\right)^{-1.75\pm0.03}, \;\;\;\; 
0.1h^{-1}<r<16h^{-1}\mathrm{Mpc}
,
\end{equation}
where $r_0\simeq6.1\pm0.2h^{-1}\mathrm{Mpc}$.
The geometry of samples in SDSS is quite close to the Las Campanas Redshift Survey \cite{shektman} and the results are very similar, but with much better resolution.

The largest data set today is a 2dF Galaxy Redshift Survey \cite{peacock01} (see fig.\ref{fig:2dF}) that consists of approximately 250,000 galaxy redshifts. Their results are:
\begin{eqnarray*}
\xi_{g}
=\left(\frac{r}{r_0}\right)^{-1.87}&\qquad\qquad 
r_0\sim(6\div10)h^{-1}\mathrm{Mpc}
,\\
\xi_{g}
=\left(\frac{r}{r_0}\right)^{-1.76}&\qquad\qquad 
r_0\sim(3\div6)h^{-1}\mathrm{Mpc}
.\\
\end{eqnarray*}
Their measurements are in agreement with previous surveys. However, having much smaller statistical errors they were able to find a slight difference in the power law exponent as well as in the correlation length for distances or redshifts, colors and types of galaxies. The result can be found at \cite{Norberg}.

\subsection{Power law clustering and fractals}

It is clear that once a correlation function is given, the density of objects around any randomly chosen member of the system is:
\begin{equation}
n(r)\propto 1+\xi(r)
.
\end{equation}
If the correlation function has a power law behaviour with exponent $\gamma$ then: 
\begin{equation}
\xi(r)\propto r^{-\gamma}
.
\end{equation}
As is the case for galaxies and clusters of galaxies, where $\gamma\simeq 1.8$, the number of objects in a given volume scales in a similar way:
\begin{equation}
N(r)\propto r^{3-\gamma}
.
\end{equation}
So for noninteger $\gamma$, the number of objects scales with a fractional power of the radius of the volume under consideration. This behaviour is typical of fractal sets.

A fractal is a set in which `mass' and `radius' are linked by a fractional power law \cite{Mandelbrot}:
\begin{equation}
M(r)\propto r^{D_F}
,
\end{equation}
where $D_F$ is the fractional or Hausdorff dimension of the set. So galaxies seem to show, at least up to scales of about 100 Mpc, a fractal distribution with $D_F\simeq1.2$. 

\begin{figure}
\begin{center}
\includegraphics[width=0.5\hsize,clip]{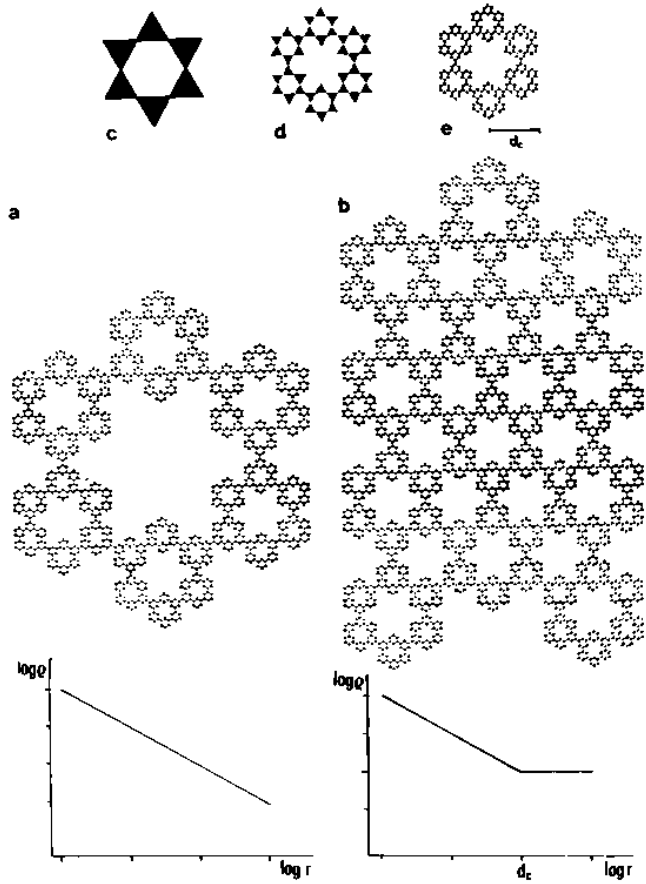}
\caption{Two fractal structures {\bf a} and {\bf b} of dimension $d\simeq 1.6$ generated by the same algorithm as {\bf c},{\bf d} and {\bf e} and endowed with the same lower cutoff, are compared and contrasted. {\bf a} has no upper cutoff, while {\bf b} has an upper cutoff at a distance $d_c$, shown in {\bf e}. In {\bf a} the density decreases with distance as $\rho\propto r^{-1.6}$, while in {\bf b}, for $r<d_c$ the distribution is self-similar and equivalent to the one in {\bf a} but for $r>d_c$ the distribution becomes homogeneous and $\rho$ stays constant with distance (from \cite{1988R}).}
\label{fig:cutoff}
\end{center}
\end{figure}

A crucial characteristic of a fractal distribution is the presence of fluctuations at all length scales and consequently the impossibility of defining an average value for the density. In a simple fractal set each observer at a matter point belonging to the set observes the same matter distribution as any other observer belonging to the set. In this sense, the fractal naturally leads to the formulation of the cosmological principle expressed by the statement (\ref{CP3}). Three very serious issues, however, arise for the compatibility of a fractal structure with that of a cosmological Friedmann model. 
(1) There must necessarily be a cutoff in the fractal distribution 
(see Fig.~\ref{fig:cutoff})in order to recover the overrall homogeneity observed at the large scales up to $z=\simeq10^3$ in the CMB. 
(2) Such a cutoff must occur homogeneously all over the Universe.
(3) The dimension of that cutoff must automatically originate as the characteristic length determined by the microphysical properties of the dark matter composing the Universe. These were the three basic thoughts which motivated the cellular model of the Universe introduced by Ruffini in the eighties \cite{1983R}-\cite{RR01}. In this model fractals arise from successive fragmentation of primordial structures, the so called `elementary cells', formed via gravitational instability in the neutrino component of the matter of the Universe. We shall further expand on this idea in the following paragraphs.

\section{Gravitational instability}

Gravitational instability is usually considered to be the basic mechanism of structure formation in the Universe (see for example \cite{kolb}). It is believed that small inhomogeneities are already present at some initial time in the early Universe. Such small perturbations will grow due to gravitational attraction, because overdense regions will accrete matter from the neighbouring regions, increasing the density contrast.

One of the simplest examples showing the process of gravitational instability is a perfect fluid model. If density distribution in a self-gravitating fluid is slightly nonuniform, i.e. small density perturbations exist, they will tend to grow. When the density contrast is small, the linear approximation can be used. The main advantage of linear theory is that perturbations on different scales evolve independently.

It is the main result of this theory that the growth of perturbations are damped by the Hubble expansion, which leads to a power law behavior for the time dependence of density perturbations. For example in the Einstein-de Sitter model which is believed to describe our Universe after recombination, perturbation amplitudes grow like $(1+z)^{-1}$. Only during the nonlinear stage with large density contrast does the evolution become faster. At the nonlinear stage, however, perturbations grow much faster, leading to the formation of gravitationally bound objects.

The theory of linear density perturbations in a homogeneous medium was first developed by Jeans \cite{jeans}-\cite{jeansbook}. His study was motivated by the intention to explain the mechanism of star formation. We describe this theory below. First, however, its range of validity,  i.e. the evolution of cosmological horizon, is discussed.

The linear perturbations in the expanding homogeneous and isotropic Friedman Universe were studied by Lifshitz \cite{lifshitz} using a relativistic treatment. Relativistic theory, however, is necessary when the scale of perturbations is greater than the horizon, or when relativistic regimes are attained in matter condensation. In the most interesting cases such as perturbations in dark matter well inside the horizon after the equivalence epoch (when the energy densities of radiation and other components are equal), it is sufficient to consider nonrelativistic theory based on Newtonian gravity. Bonnor \cite{bonnor} (see also \cite{heath}) was the first to study evolution of spherically symmetric perturbations in Newtonian cosmology.

The theory of linear density perturbations in the Newtonian treatment is developed in detail in various textbooks, 
see e.g. \cite{weinberg}-\cite{zeldovich}.

\subsection{Horizon scale and mass evolution}

The Newtonian treatment is only applicable on scales smaller than the horizon scale $\lambda_H=cH^{-1}$. The associated mass scale, defined as the mass contained within a sphere of radius $\lambda_H/2$, where $H$ is the Hubble parameter, is given by
\begin{equation}
\label{MH}
M_H=\frac{4}{3}\pi \rho\left(\frac{\lambda_H}{2}\right)^3
.
\end{equation}

Beyond this scale events are causally disconnected and thus any correlation breaks down outside the horizon. Thus structures cannot form on scales larger than $\lambda_H$. $M_H$ monotonically increases with time because the distance that light travels increases with time. There are several regimes, separated by the moment of equivalence in energy densities of radiation and nonrelativistic matter:
\begin{equation}
\label{MHevol}
M_H\propto
\left\{
\begin{array}{ll}
a^3 & z>z_{eq} \\
a^{3/2} & z<z_{eq}
.
\end{array}
\right.
\end{equation}

Today the horizon scale is approximately 3000Mpc, which corresponds to a mass scale $M\sim10^{22}M_\odot$ for an $\Omega=1$ Universe. At recombination the total mass inside the horizon was then approximately $(1/z_{rec})^{-3/2}\simeq10^{17}M_\odot$, where $M_\odot=2\,10^{30}\,$kg is a solar mass.

\subsection{Self-gravitating ideal fluid: linear theory}

\subsubsection{Fluid equations and background solutions}

Consider a perfect fluid with density $\rho$ and pressure $p$ in Euclidean space with a Cartesian ("physical") coordinate system $r_i$\footnote{Greek indices denote comoving coordinates, Latin indices denote physical coordinates, both take the values 1,2,3; the Einstein summation rule is adopted.}. The fluid has a velocity field $v_i$; the gravitational potential $\Phi$ is induced by the mass density $\rho$ distribution. All these quantities are related through the continuity, Euler and Poisson equations respectively. For a nonrelativistic fluid, i.e. for a fluid with $p<<\rho c^2$, they read \cite{peebles} \cite{weinberg} \cite{padmanabhan}  \cite{raycha}:
\begin{equation}
\label{continuity}
\frac{\partial \rho}{\partial t}+\partial_i(\rho v_i)=0
,
\end{equation}
\begin{equation}
\label{euler}
\frac{\partial v_i}{\partial t}+v_j \partial_j v_i +\frac{1}{\rho}\partial_i p+\partial_i \Phi=0
,
\end{equation}
\begin{equation}
\label{poisson}
\partial^2\Phi-4\pi G\rho=0
,
\end{equation}
where $\partial^2=\partial_i \partial_i$. A cosmologically important solution of equations (\ref{continuity}-\ref{poisson}) is the one describing a spatially uniform fluid with zero pressure (`dust') on an expanding background \cite{weinberg} \cite{raycha}:
\begin{equation}
\label{hubble}
v^0_i=H(t)r_i
,
\end{equation}
\begin{equation}
\label{density}
\frac{d\rho_0}{dt}+3H\rho_0=0
,
\end{equation}
\begin{equation}
\label{pressure}
p_0=p_0(t)
,
\end{equation}
\begin{equation}
\label{potential}
\Phi_0=\frac{2}{3}\pi G\rho_0 r^2
,
\end{equation}
\begin{equation}
\label{friedman}
\frac{dH}{dt}+H^2=-\frac{4}{3}\pi G\rho_0
,
\end{equation}
where $r^2=r_i r_i$ and all quantities depend only on time.

In a comoving coordinate system, with 
\begin{equation}
\label{comoving}
r_\alpha=a(t)x_\alpha
,
\end{equation}
where $a(t)$ is the scale factor, the relation between coordinate differences $\Delta r_i$ and $\Delta x_\alpha$ is
\begin{equation}
\label{coorddiff}
\frac{d\Delta r_\alpha}{dt}
=a\frac{d\Delta x_\alpha}{dt}+\frac{da}{dt}\Delta x_\alpha
=a\frac{d\Delta x_\alpha}{dt}+H(t)\Delta r_\alpha
,
\end{equation}
where 
\begin{equation}
\label{H0}
H(t)=\frac{1}{a}\frac{da}{dt}
.
\end{equation}

Correspondingly, for velocity fields we have:
\begin{equation}
\label{velocity}
v_\alpha(r_\beta,t)
=u_\alpha(x_\beta,t)+H r_\alpha=u_\alpha(x_\beta,t)+v^0_\alpha
.
\end{equation}

Thus the solution (\ref{hubble}-\ref{friedman}) represents a uniform distribution of the fluid with zero peculiar velocity $u^0_\alpha=0$ and zero pressure $p_0=0$.
Pressure and density are linked through the equation of state $p=p(\rho)$. The three equations (\ref{continuity}-\ref{poisson}) together with the equation of state are a complete set, allowing one to study the temporal evolution of the density and velocity distributions as well as of the pressure and gravitational potential.

\subsubsection{Perturbed quantities}

As well known, solutions (\ref{hubble}-\ref{friedman}) represent an isotropic and homogeneous distribution of matter. In order to study density perturbations in the linear approximation assume that
\begin{equation}
\label{densityperturb}
\rho(r_i,t)=\rho_0(t)\left[1+\delta(r_i,t)\right]
,
\end{equation}
\begin{equation}
\label{velocityperturb}
v_i(r_j,t)=v^0_i(r_j,t)+\delta v_i(r_j,t)
,
\end{equation}
\begin{equation}
\label{potentialperturb}
\Phi(r_i,t)=\Phi_0(r_i,t)+\delta\Phi(r_i,t)
,
\end{equation}
\begin{equation}
\label{pressureperturb}
p(r_i,t)=\delta p(r_i,t)
,
\end{equation}
where $\delta\equiv\frac{\rho-\rho_0}{\rho_0}$. Here all perturbed quantities $\delta$, $\delta v_i$, $\delta p$ and $\delta \Phi$ are assumed to be much smaller than the background quantities. All zero order values are given by (\ref{hubble}-\ref{friedman}). Assume also, that spatial together with temporal derivatives of perturbed quantities have the same order of magnitude as the quantities themselves.

Note that it is not necessary for the smallness condition on the perturbed quantities to hold over all of space. In particular,
there could be a region in space where  $|\delta v_i|>|v^0_i|$ \cite{meszaros}. In this case the standard linearization procedure leads to different perturbation equations and consequently to different solutions representing a time dependent density contrast $\delta(r_i,t)$.

\subsubsection{Linearized perturbations equations}

Rewriting (\ref{continuity}-\ref{poisson}) in comoving coordinates:
\begin{equation}
\label{comcontinuity}
\frac{\partial \rho}{\partial t} 
+3H\rho+\frac{1}{a}\rho\partial_\alpha u_\alpha
+\frac{1}{a}u_\alpha \partial_\alpha\rho=0
,
\end{equation}
\begin{equation}
\label{comeuler}
\frac{d^2a}{dt^2}x_\alpha
+\frac{\partial u_\alpha}{\partial t} +Hu_\alpha+\frac{1}{a}u_\beta\partial_\beta u_\alpha+\frac{1}{a\rho}\partial_\alpha p
+\frac{1}{a}\partial_\alpha \Phi=0
,
\end{equation}
\begin{equation}
\label{compoisson}
\partial^2\Phi-4\pi Ga^2\rho=0
.
\end{equation}

Here all quantities, except for $H$, depend on the comoving coordinates\footnote{If one also assumes that Hubble parameter can be disturbed (have spatial dependence) then the system of equations becomes overdefined. There is another approach \cite{EB}, however, where $\partial_\alpha\delta$ and $\partial_\alpha H$ are taken as independent variables in order to study density perturbations.} $x_\alpha$ and the time $t$. 
Eqs.~(\ref{comcontinuity}-\ref{compoisson}) written in physical coordinates can be found in \cite{meszaros} for example. One arrives at the above results from (\ref{continuity}-\ref{poisson}) by using the transformation laws 
$(\partial/ \partial t)_{phys} =(\partial/ \partial t)_{com}
-Hx_\alpha\partial_\alpha$ and $(\partial_\alpha)_{phys}=(1/a)(\partial_\alpha)_{com}$.

In order to obtain equations for the density contrast $\delta$ in the linear approximation substitute (\ref{densityperturb}-\ref{pressureperturb}) into equations (\ref{comcontinuity}-\ref{compoisson}). Taking into account that spatial as well as temporal derivatives of perturbed quantities have the same order of smallness as the perturbed quantities themselves and using (\ref{hubble}-\ref{friedman}), the perturbation equations read
\begin{equation}
\label{delta}
\frac{\partial \delta}{\partial t}+\frac{1}{a}\partial_\alpha\delta u_\alpha=0
,
\end{equation}
\begin{equation}
\label{deltau}
\frac{\partial\delta u_\alpha}{\partial t}+H\delta u_\alpha+\frac{1}{a}\partial_\alpha \delta p+\frac{1}{a}\partial_\alpha \delta \Phi=0
,
\end{equation}
\begin{equation}
\label{deltaphi}
\partial^2\delta \Phi-4\pi Ga^2\rho_0 \delta=0
,
\end{equation}
where $\delta u_\alpha(x_\beta,t)$ is first order quantity, because the unperturbed value is $u^0_\alpha(x_\beta,t)=0$.

The simplest way to find the equation governing density perturbations is to take the time derivative of Eq.~(\ref{delta}) and use the divergence of Eq.~(\ref{deltau}) together with Eq.~(\ref{deltaphi}).
After rather long calculations one finds the final expression:
\begin{equation}
\label{d2delta}
\frac{\partial^2 \delta}{\partial t^2}
+2H\frac{\partial \delta}{\partial t}
-\frac{v_s^2}{a^2}\partial^2 \delta-4\pi G \rho_0 \delta=0
,
\end{equation}
where the relation
\begin{equation}
\label{soundspeed}
v_s^2=\frac{dp}{d\rho}
\end{equation}
is assumed.

\subsubsection{The Jeans criterion}

Eq.~(\ref{d2delta}) governs the dynamics of density perturbations. It is a wave-like second order partial differential equation. Thus it is natural to introduce the Fourier transform
\begin{equation}
\label{furier}
\delta=\sum_{k}h(t)e^{ik_\alpha x_\alpha}
\end{equation}
in order to split perturbations on different scales.

Eq.~(\ref{d2delta}) can be rewritten in $k$-space, taking into account that $\partial_\alpha \delta \rightarrow ik_\alpha h$:
\begin{equation}
\label{ke}
\frac{d^2 h}{dt^2}
=-2H\frac{dh}{dt}+(4\pi G \rho_0-\frac{v_s^2k^2}{a^2})h
,
\end{equation}
where $k_\alpha$ is the comoving wavevector and $k=\sqrt{k_\alpha k_\alpha}$ is the corresponding wavenumber. The comoving wavelength of the perturbative mode is given by $l=2\pi/k$, while the proper (physical) wavelength is simply $\lambda=al$.

The Jeans criterion (\ref{ke}) is governed by the wavelength
\begin{equation}
\label{jeans}
\lambda_J=v_s\sqrt{\frac{\pi}{G \rho_0}}
,
\end{equation}
where $\lambda_J$ separates gravitationally stable scales from unstable ones. Fluctuations on scales well above $\lambda_J$ grow via gravitational instability, while on scales smaller than $\lambda_J$ the pressure overwhelms gravity and perturbations do not grow.

The first term on the right-hand side of (\ref{ke}) comes from the general expansion. In the static world initially considered by Jeans, such a term is absent, leading to the exponential growth of perturbations. In expanding space perturbations grow with time according to a power law.

A very important quantity usually associated with the Jeans length  is the Jeans mass (\ref{jeans})
\begin{equation}
\label{JeansMass}
M_J=\frac{4}{3}\pi \rho \left(\frac{\lambda_J}{2}\right)^3
,
\end{equation}
defined as the mass contained within a sphere of radius $\lambda_J/2$, where $\rho$ is density of the perturbed component.

\subsubsection{Multi-component system}

Perturbations for a given mode in a single component evolve according to (\ref{ke}). When several components such as Cold Dark Matter (CDM), Hot Dark Matter (HDM), baryons and radiation are present simultaneously, it is possible to generalize (\ref{d2delta}). Assuming gravitational interaction between components only, we arrive at
\begin{equation}
\label{multi}
\frac{d^2 h_i}{dt^2}
=-2H\frac{dh_i}{dt}
+(4\pi G \rho_0 \sum_{j}\epsilon_j h_j-\frac{(v_s^2)_i k^2}{a^2}h_i)
,
\end{equation}
where the index $i$ refers to the component under consideration, the sum is over all components and $\epsilon_i=\rho_i/ \sum_j \rho_j$. Notice that any smoothly distributed component (like the cosmological constant) does not contibute to the right-hand side of (\ref{multi}).

\subsection{Applications}

Some important cases of matter content for the Unverse will be considered below. First we discuss perturbation dynamics in the dominant nonrelativistic component (baryonic or not). Second example is dark matter perturbations in the presence of a dominant radiation component.

\subsubsection{Einstein-de Sitter Universe}

First consider the dust dominated $\Omega=1$ Universe. This condition ($\Omega$ is the density parameter) corresponds to a flat-type cosmological model, namely the Friedman solution of Einstein's General Relativity equations with curvature parameter $k=0$. This model is thought to provide a good description of our Universe after recombination.
To zero order $a\sim t^{2/3}, \;\; H=2/3t$ and $\rho_0=1/6\pi G t^2$. For perturbations well inside the horizon we have
\begin{equation}
\label{EdS}
t^2 \frac{d^2h}{dt^2}
+\frac{4}{3}t\frac{dh}{dt}
-\frac{2}{3}\left[1-\left(\frac{\lambda_J}{\lambda}\right)^2\right]h=0
.
\end{equation}

For modes well inside the horizon and still larger than the Jeans length the solution is
\begin{equation}
\label{EdSsol}
h(k,t)=h_1(k) \left(\frac{t}{t_0}\right)^{2/3} 
+h_2(k) \left(\frac{t}{t_0}\right)^{-1}
.
\end{equation}

As expected there are two solutions, one growing and one decaying. At late time, however, only the growing mode is important. Perturbations evolve proportionally to the scale factor or as $(1+z)^{-1}$, where $z$ is the redshift defined by:
\begin{equation}
\label{redshift}
1+z=\frac{a_0}{a(t)}
,
\end{equation}
and $a_0$ denotes the value of a scale factor today.

Perturbations on scales smaller than the Jeans length cease to grow and oscillate with time.

\subsubsection{Mixture of radiation and dark matter}\label{ssec:MRD}

Consider the radiation dominated Universe where $a\sim t^{1/2}$ and $H=1/2t$. The second component to be considered is a collisionless dark matter with $v_{DM}=0$. We still can use the Newtonian treatment on scales much smaller than the horizon size.

Since the small scale photon distribution is smooth and the energy density is dominated by the radiation, the equation governing dark matter instability reduces to
\begin{equation}
\label{R+DM}
t \frac{d^2h_{DM}}{dt^2}+\frac{dh_{DM}}{dt}=0
.
\end{equation}
This has the solution
\begin{equation}
\label{R+DMsol}
h_{DM}(k,t)=h_1(k)\log\left(\frac{t}{t_0}\right)+h_2(k)
.
\end{equation}
Perturbations in dark matter component inside the horizon experience a slow logarithmic growth. This is the well known Meszaros effect \cite{oldmeszaros}.

\subsection{Initial spectrum of perturbations}

In the first example we have shown that perturbations on scales between horizon size and Jeans length $\lambda_J\ll\lambda\ll\lambda_H$ grow as $\delta\propto(1+z)^{-1}$. In order to study perturbation dynamics one 
also needs to know the initial values of the perturbations at some moment in early Universe.

One great possibility is provided by the CBR anisotropy measurements, because density inhomogeneities when photons were coupled to baryons can be extracted from temperature fluctuations observed in the CBR. Since $\frac{\delta T}{T}\simeq 10^{-5}$ it is usually assumed that at the moment of recombination $\delta\simeq 10^{-4}$ \cite{kolb}.

Perturbation amplitudes on different scales are usually represented by a power spectrum $P(k)$, which is the Fourier transform of a previously introduced correlation function \cite{peacockbook}
\begin{equation}
\xi(r)=\frac{V}{(2\pi)^3}\int P(k)\frac{\sin kr}{kr}4\pi k^2dk
.
\end{equation}
There is no evidence that the initial spectrum contained any preferred scale, so it should be a featureless power law
\begin{equation}
\label{insp}
P(k)\propto k^n
,
\end{equation}
where the index $n$ governs the balance between perturbation amplitudes on large and small scales. The value $n=0$ corresponds to white noise that has the same amplitude for every mass scale. The value $n=1$ corresponds to a so-called Harrison-Zel'dovich scale invariant spectrum. Term `scale invariance` means that perturbations had the same amplitude at the moment of horizon crossing.

\subsection{Damping of perturbations}

In addition to the Jeans scale some other cosmologically important scales appear in the theory of structure formation. They are related to physical processes that cannot be described within the perfect fluid approximation. However, fortunately such processes take place on limited scale intervals and outside such intervals the fluid description is still possible. We will discuss some dissipative effects, such as collisional damping of baryonic perturbations and free streaming of collisionless light particles.

\subsubsection{Silk damping}

Close to recombination the coupling between photons and baryons makes it possible for the former to erase perturbations of the latter. This is because at that time the free mean path of photons becomes larger, so they can travel from overdense into underdense regions dragging baryons with them, thus smoothing inhomogeneities in the primeval plasma. This effect was discovered by Silk \cite{Silk}. The physical scale associated with it is \cite{kolb}
\begin{equation}
l_S \simeq 3.5(\Omega h^2)^{-3/4}\,\mathrm{Mpc}
,
\end{equation}
which gives a mass scale
\begin{equation}
M_S \simeq 6.2 \;\; 10^{12}(\Omega h^2)^{-5/4}\,\mathrm{M}_{\odot}
.
\end{equation}
This scale is close to the mass of a typical galaxy $10^{11}M_{\odot}$. However, Silk damping only affects baryonic perturbations. Moreover, it is important only around recombination when the coupling is still sufficiently strong to make photons drag baryons with them.

\subsubsection{Free streaming}\label{ssec:FS}

Another dissipative process is Landau damping or free streaming that originates from the free motion of collisionless particles on small scales. They can travel far if the velocity dispersion is large. This is important after particles decouple from the plasma and until they become nonrelativistic.

The discovery of free streaming was a dramatic moment for structure formation scenarios based upon Hot Dark Matter (HDM) models \cite{FS}. 
The name Hot Dark Matter means that particles composing such matter were ultrarelativistic at equivalence. Thus their velocity dispersion was near the speed of light $c$.
The maximum distance scale travelled by collisionless particles from decoupling can be estimated to be \cite{padmanabhan}
\begin{equation}
l_{FS} \simeq 0.5(\frac{m_{DM}}{1\,\mathrm{keV}})^{-4/3}
             (\Omega_{DM} h^2)^{1/3}\,\mathrm{Mpc}
,
\end{equation}
and a corresponding mass scale is of  the order of superclusters of galaxies or even larger if the particle mass is $m_{DM}<30\,$eV.

Cold thermal relics which compose the Cold Dark Matter (CDM) are slow enough so that free streaming can be neglected on cosmologically important scales. Therefore Landau damping affects only light particles like neutrinos with $m_\nu\sim10\,$eV, and in the Universe dominated by HDM, all perturbations on scales smaller than superclusters of galaxies are erased. At the same time, after particles become nonrelativistic at $z_{nr}$, their velocity dispersion becomes small enough to make free streaming negligible.

\subsection{Structure formation at late times}

\subsubsection{Nonlinear clustering}

In previous sections we dealt with the linear evolution of cosmological perturbations. From long after recombination and even earlier until almost recent times, such a treatment is  justified to describe the growth of inhomogeneity in the Universe because the condition $\delta\ll 1$ is sutisfied. In recent times, say at $z\sim10$, the nonlinear behaviour of perturbations becomes important. During the nonlinear stage gravitationally bound objects such as galaxies form. Nonlinear evolution is a rapid process, where not only gravitational effects are important. In particular during the formation of galaxies, various dissipation and relaxation processes take place \cite{peacockbook}.

We will not go into the details of galaxy formation here. The LSS formation is the subject of this section. Here the main interaction remains gravity. However, the theoretical description based on linear equations for an ideal fluid becomes inadequate.

Usually N-body simulations are employed to study nonlinear clustering \cite{peacockbook}. However, some useful simplified models are still possible even in the nonlinear stage, because numerical simulations provide limited physical insight into the physics of gravitational clustering. Among nonlinear approximations, the most famous are the Zel'dovich approximation \cite{Zel} and the spherical collapse model (see for example \cite{peebles}).
The key point in the Zel'dovich model is that during collapse in an almost spherically symmetric overdense region, gravitational interaction amplifies asymmetry. Therefore, the final structure acquires a preferred direction and the final collapsed body will look like a `pancake'.
In the spherical collapse model, on the other hand, it is assumed that spherical symmetry remains valid during the entire period of collapse. It allows the splitting of the spherical overdense region into concentric shells and the evolution of each shell can be studied separately, which sufficiently simplifies the problem. This model will be described in detail in the next chapter.

\subsubsection{Structure formation scenarios}

Historically two different pictures of structure formation were considered, namely the HDM (see \cite{pHDM}) and CDM models (see \cite{pCDM}). We will discuss them briefly below.

\subsubsection{HDM models}

The neutrino dominated Universe with $m_\nu\sim 10\,$eV is a typical HDM model. The HDM model is associated with so-called ``top-down" scenario, where structures form on large scales first. This is so because the Jeans mass for HDM is of the order of the supercluster mass or even higher. At the same time, free streaming erases perturbations on smaller scales. Thus only when perturbations reach the nonlinear regime on large scales can  they induce fragmentation on smaller scales.

Usually it is assumed that large scale perturbations become nonspherical according to the Zel'dovich model and thus the LSS look like a ``net" of density condensations separated by huge voids. Simulations agree with such a picture.

The HDM model is in good agreement with observational data on scales larger than $10\,$Mpc. On smaller scales, however, HDM simulations can agree with observed the correlation function of galaxies only if the epoch of pancaking takes place at $z\simeq 1$ or less, which is too late, because we can see galaxies and quasars with much greater $z$.

The crucial cosmological property of HDM with neutrinos, as was mentioned above, is the damping of perturbations on small scales due to free streaming. Neutrino dominated models ($\Omega_\nu\sim1$) alone could not describe the real Universe because on scales smaller than $\sim100\,$Mpc no structure appears at all.

One important prediction of HDM models with neutrinos is the existence of large smooth halos around galaxies. At the end of collapse, during formation of galaxies, the baryonic component can dissipate its energy via collisions, but the neutrino component cannot. Thus neutrinos remain less condensed than baryons, forming large galactic halos.

\subsubsection{CDM models}

CDM models do not have trouble with free streaming, because their particles have negligible velocities at decoupling. Moreover, the Jeans mass for typical CDM model lays well below $10^6M_\odot$. Thus, perturbations start to develop on small scales simultaneously with perturbations on large scales.

In the CDM models the important feature is the weak growth experienced by perturbations between horizon crossing and equivalence (see \S\ref{ssec:MRD}). This means that the density contrast increases when we move to smaller scales, or that the perturbations spectrum has more small-scale power.

After collapse the first structures in CDM models virialize  through violent relaxation \cite{LBS} into gravitationally bound objects that form galactic halos. Structures form in a self-similar manner from small to large scales, in other words according to a `bottom-up' scenario.

Pure CDM models, however, fail to predict the observed correlation function for galaxies on large scales. If one wants to retain the CDM hypothesis, the simplest way is to reduce the matter density. This shifts matter-radiation equivalence to a later epoch, resulting in redistibution of power in the spectrum of perturbations in favour of larger scales.

Today the $\Lambda$CDM model with $\Omega_{tot}=1$ and $\Omega_\Lambda=0.7$ is considered to be the best fit to the full set of observational data (see E.W. Kolb, in these proceedings \cite{kolbBr}).

\section{Neutrinos and structure formation}

In previous chapter we described the evolution of perturbations and we saw that the nature of dark matter particles is crucial to determining the way structure formation develops. In spite of the fact that a lot of candidates for CDM particles exist, there is presently no experimental evidence of such particles. On the other hand, neutrinos are the only candidates for DM known to exist. 

`Light' neutrinos ($m_\nu\ll1$MeV) \cite{DBook}, namely neutrinos that decouple while still in their ultrarelativistic regime (see below), may provide a significant contribution to the energy density of the Universe ($\Omega_\nu\sim 1$). Models with light neutrinos were extensively studied in the eighties; a large literature exists on this subject \cite{Mnu}.

\begin{figure}[t]
\begin{center}
\includegraphics{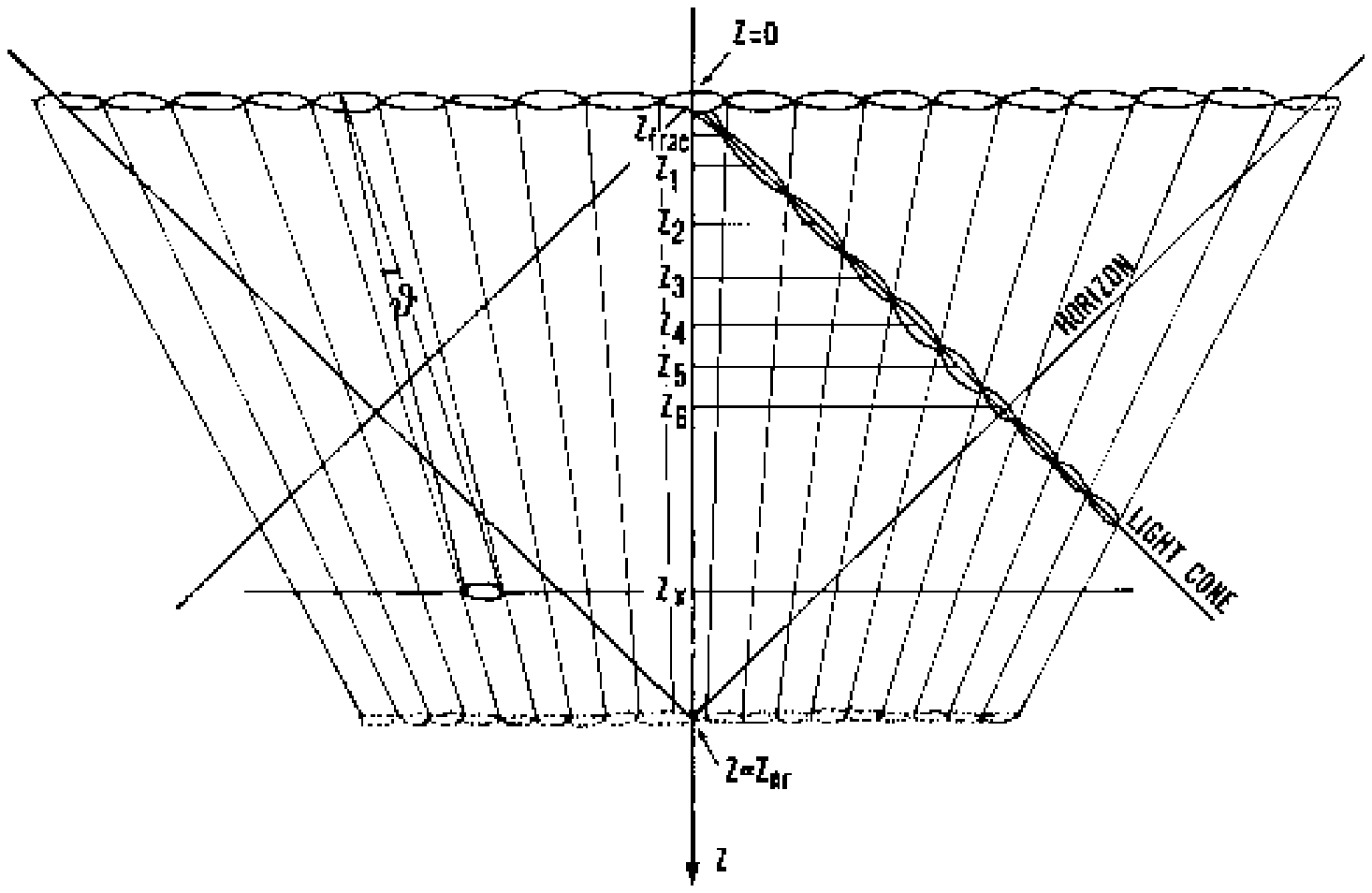}
\caption{Qualitative drawing of the evolution of the cellular structure of the Universe. The vertical axis represents decreasing values of the redshift. $z=0$ represents the cosmological observer today, embedded in his own `elementary cell`; $z_1\,z_2,...,\,z_n$ are the redshifts of the centers of the successive `elementary cells` seen by this observer. $z=z_\gamma$ represents the surface of last scattering (decoupling) of the CBR. $\theta$ is the angular size subtended today by an `elementary cell` at the time of decoupling. Finally, $z=z_{nr}$ is the time at which the cells formed via gravitational instability.}
\label{fig:cell}
\end{center}
\end{figure}

The key prediction of the cosmological model with neutrinos is a cellular structure on large scales (see Fig.~\ref{fig:2dF}).  The qualitative drawing of cellular structure of the Universe is represented in Fig.~\ref{fig:cell}.
Ruffini and collaborators have studied such models with particular attention to the problem of clustering on large scales and its relation to the fractal distribution of matter. In the following, we will outline some of these ideas.

\subsection{Neutrino decoupling}

The cosmological evolution of a gas of particles can be split into two very different regimes. At early times, the particles are in thermal equilibrium with the cosmological plasma; this corresponds to the situation in which the rate $\Gamma=<\sigma v n>$ of the reactions supposed to mantain the equilibrium (such as $\nu_e+\bar\nu_e\leftrightarrow e^++e^-\leftrightarrow 2\gamma$ in the case of electronic neutrinos) is much greater than the expansion rate, given by the Hubble parameter. The gas then evolves through a sequence of thermodynamic equilibrium states, described by the usual Fermi-Dirac statistics:
\begin{equation}
f(p)=\frac{1}{\exp\left[(E(p)-\mu)/k_BT\right]+1}
,
\end{equation}
where $p$, $\mu$ and $T$ are the momentum, chemical potential and temperature of neutrinos respectively, and $k_B$ is the Boltzmann constant.

However, as the Universe expands and cools, the collision rate $\Gamma$ becomes smaller than the expansion rate; this means that the mean free path is greater than the Hubble radius, so we can consider the gas to be expanding without collisions. It is customary to describe the transition beetween the two regimes by saying that the gas has decoupled from the cosmological plasma.

\subsubsection{The redshifted statistics}

Since in a spatially homogeneous and isotropic Universe described by the Robertson-Walker metric, the product of the three-momentum $p(t)$ of a free particle times the scale factor $a(t)$ is a constant of the motion:
\begin{equation}
p(t)\cdot a(t)=\mathrm{const}
,
\end{equation}
each particle in the gas changes its momentum according to this relation. This fact, together with Liouville's theorem, implies that the distribuition function after the decoupling time $t_{d}$ (defined as the time at which $\Gamma=H$) is given by \cite{1983R}:
\begin{equation}
f(p, t>t_{d})=f\left(\frac{a(t)}{a_{d}}p,\,t_{d}\right)
=
\frac{1}{\exp\left[(E\left(\frac{a(t)}{a_{d}}p\right)-\mu_{d})/k_B T_{d}\right]+1}
,
\label{Eq RS Stat}
\end{equation}
where the subscript $d$ denotes quantities evaluated at the decoupling time.

Now we turn our attention to the special case of neutrinos with $m_\nu\lesssim 10\,$eV. The ratio $\Gamma/H$, as a function of the cosmological temperature, can be evaluated using quantum field theory \cite{kolb}
\begin{equation}
\frac{\Gamma}{H}\simeq\left(\frac{T}{1\,\mathrm{MeV}}\right)^3
\end{equation}
as long as $T\gg m$.
Therefore, neutrinos decouple from the cosmological plasma when $T=T_d\simeq1\,\mathrm{MeV}$. Since $kT_d\gg m c^2$, many of the particles satisfy $pc\gg mc^2$ and then, when performing the integration over the distribution function (\ref{Eq RS Stat}), we can safely approximate:
\begin{equation}
f(p, t>t_{d})=f\left(\frac{a(t)}{a_{d}}p,\,t_{d}\right)\simeq
\frac{1}{\exp\left[\left(\displaystyle\frac{a(t)}{a_{d}}pc-\mu_{d}\right)/k_B T_{d}\right]+1}
,
\label{Eq RS Stat Appr}
\end{equation}
since the tail of the distribution function for which $mc^2\gg pc$ gives little contribution. 

In the following, we compute the mean value of physical quantities over this distribution. It will be useful to consider two limiting regimes, namely the nonrelativistic one and the ultrarelativistic one. They correspond to two approximations for the single particle energy \cite{1983R}:
\begin{eqnarray}
\begin{array}{ll}
E\simeq mc^2&\qquad kT\ll mc^2
,
\qquad\textrm{NR}
,\\
E\simeq pc&\qquad kT\gg mc^2
,
\qquad\textrm{UR}
.
\end{array}
\end{eqnarray}
We emphasize the fact that this substitution must be performed only in the function to be integrated, and not in the distribution function. The approximation (\ref{Eq RS Stat Appr}) depends only on the fact that the particles are ultrarelativistic at the time of decoupling, and then it is valid even when $kT\ll mc^2$.

Then, with a suitable substitution of variables, all the relevant integrals can be recast into a very simple dimensionless form:
\begin{equation}
I_n(\xi)\equiv
\int_0^\infty{\frac{y^ndy}{\exp\left[\left(y-\xi\right)\right]+1}}
,
\end{equation}
where $\xi\equiv\mu_d/kT_d$ is the dimensionless chemical potential or degeneracy parameter.
These integrals can be expressed using Riemann zeta and related functions \cite{LL}.

\subsubsection{Energy density of neutrinos}

The present density parameter of neutrinos can be easily evaluated using the method outlined in the previous section. The energy density is given by:
\begin{equation}
\rho_{\nu+\bar\nu}(t_0)=\frac{g}{h_{P}^3}\int_0^\infty E(p)f(p, t_0)\,d^3p
,
\end{equation}
where $g$ is the number of helicity states and $h_{P}$ is Planc's constant.
By normalization with  respect to the critical density $\rho_c=1.054\,h^2\cdot10^4 \frac{\mathrm{eV}}{\mathrm{cm^3}}$, we obtain \cite{1986R} \cite{1988R}:
\begin{equation}
\Omega_{\nu+\bar\nu}h^2\simeq 1.10\cdot10^{-1} \,g\,\frac{m}{10\,\mathrm{eV}}\,A(\xi)
,
\end{equation}
where $A(\xi)$ is defined as follows
\begin{equation}
A(\xi)\equiv\frac{I_2(\xi)+I_2(-\xi)}{2I_2(0)}
\\
=\frac{1}{4\eta(3)}\left[\frac13\,|\xi|^3+4\eta(2)|\xi|
 +4\sum_{k=1}^\infty(-1)^{k+1}\,\displaystyle\frac{e^{-k|\xi|}}{k^3}\right]
,
\end{equation}
and $\eta(n)$ is the Riemann eta function of index $n$.

The term $I_2(-\xi)$ appears because we have to take into consideration the presence of antiparticles, for which the relation $\xi_{\bar\nu}=-\xi_\nu$ holds. This result follows from the fact that if we consider a reaction such as
\begin{equation}
\nu+\bar \nu\longleftrightarrow ... \longleftrightarrow \gamma+\gamma
,
\end{equation}
then since the chemical potentials of the initial and final states have to be equal and the chemical potential of the latter is equal to zero, it follows that $\xi_{\bar \nu}=-\xi_\nu$.

\subsubsection{Recent constraints on the neutrino mass $m_\nu$ and degeneracy parameter $\xi_\nu$}

We discuss below very briefly recent bounds on the chemical potential and mass of neutrinos \cite{GM} that can be used to compute the recent bound on $\Omega_{\nu+\bar\nu}$.

\paragraph{Neutrino mass}

A recent laboratory limit on the electron neutrino mass comes from tritium $\beta$ decay \cite{mLab}. These data give limits 
\begin{equation}
\label{m}
m_{\nu_e}<2.5\,\mathrm{eV}
.
\end{equation}

At the same time, no direct measurements or constraints on muonic and tauonic neutrino masses exist. Moreover, it is still unknown whether neutrinos are Majorana or Dirac particles.
Very recent data from neutrinoless double $\beta$ decay \cite{meLab} give also lower bound on Majorana mass:
\begin{equation}
\label{eemass}
(0.05\leq m_{\nu_{ee}}\leq 0.86)\,\mathrm{eV}
.
\end{equation}

\paragraph{Chemical potential}

The first constraints on the neutrino degeneracy parameter from cosmological nucleosynthesis were obtained in \cite{fBBN}. It was shown later \cite{1991R} that a small value of $\xi_e$ coupled with large values of $|\xi_{\mu,\tau}|$  can lead to cosmological nucleosynthesis abundances which are consistent with observations. It is found in particular that
\begin{equation} 
0\leq\xi_e\lesssim1.5
,
\end{equation}
with the additional constraint $F(\xi_\mu)+F(\xi_\tau)\approx F(10\xi_e)$, where $F(\xi)\equiv\xi^2+\xi^4/2\pi^2$. In particular this implies $|\xi_{\mu,\,\tau}|\lesssim10\xi_e$.

Recent data both from cosmological nucleosynthesis and CMBR \cite{eBBNCMBR} strongly constrain neutrino degeneracy parameters. Orito et al.~\cite{OritoBBN} give surprisingly wide constraints, $\xi_e<1.4$ and $|\xi_{\mu,\tau}|<40$. Other papers give essentially stronger constraints using additional assumptions \cite{KnBBN}--\cite{eBBNCMBR} 
\begin{equation}
\begin{array}{l}
\xi_e<0.3\\
|\xi_{\mu,\tau}|<2.6
.
\end{array}
\end{equation}

\paragraph{Neutrino oscillations}

When one consider different chemical potentials for all neutrino flavors at the epoch prior to cosmological nucleosynthesis, neutrino oscillations equalize chemical potentials \cite{eqmu} if there is enough time for the relaxation process \cite{AbBBN}. On the basis of large mixing angle solution of the solar neutrino problem which is favored by recent data \cite{SNO}, cosmological nucleosynthesis considerations constrain the degeneracy parameters of all neutrino flavors \cite{nuDeg}:
\begin{equation}
|\xi|\leq 0.07
. 
\end{equation}

However, the situation when flavor equilibrium is not achieved before cosmological nucleosynthesis is also possible. Thus in the following we consider quite high values of the degeneracy parameter and assume its value positive without loss of generality.

The main result that comes from considering oscillations  is that masses of different neutrino species are nearly equal: $m_{\nu_e}\simeq m_{\nu_\mu}\simeq m_{\nu_\tau}$.

\vspace{1cm}

One can see that quite high values of the neutrino energy density are still possible if we assume the recent constraints discussed above. In particular, if one consider two Dirac neutrino flavors with equal masses and chemical potentials ($\nu_e$ gives very small contribution to $\Omega_\nu$), $\xi_{\mu}=\xi_{\tau}\lesssim 2.6$ \cite{Orito2BBN} and $m_{\nu_{\mu}}=m_{\nu_{\tau}} \leq 2.5\,$eV \cite{mLab}, one gets the upper bound
\begin{equation}
\label{uub}
\Omega_{\nu+\bar{\nu}}\leq 0.45
.
\end{equation}

\subsubsection{The Jeans mass of neutrinos}

In neutrino dominated Universe the first possible structure occurs when
these particles become nonrelativistic, since at earlier times free streaming erases all perturbations.  At this epoch the cosmological redshift has the value \cite{1988R}
\begin{equation}
\label{rs}
1+z_{nr}=1.698 \; 
10^4\left(\frac{m_{\nu}}{10eV}\right)A(\xi)^{\frac{1}{2}}B(\xi)^{-\frac{1}{2}}
,
\end{equation}
where
\begin{equation}
\label{Beta}
B(\xi)\equiv\frac{I_3(\xi)+I_3(-\xi)}{I_3(0)}
=\frac{1}{48\eta(5)}\left[\frac{1}{5}\xi^5+8\eta(2)\xi^3+48\eta(4)\xi
+48\sum_{n=1}^{\infty}(-1)^{n+1}\frac{e^{-n\xi}}{n^5}\right]
.
\end{equation}

The basic mechanism of fragmentation of the initial inhomogeneities in an
expanding Universe is the Jeans instability described in the previous section.
However, in the calculation of the Jeans length of nonrelativistic collisionless neutrinos, we cannot use the velocity of sound obtained from the classical formula (\ref{soundspeed}). In fact, since the particles are collisionless, their effective pressure is zero and this would lead to a vanishing Jeans length, meaning that even the smallest perturbation would be unstable. This is not the case since in the absence of pressure, another mechanism works against gravitational collapse, namely the free streaming of particles (see \S \ref{ssec:FS}). The characteristic velocity associated with this process is simply the dispersion velocity $\sqrt{<v^2>/3}$, where the factor 3 comes from averaging over spatial directions.
Thus, we have to make the substitution $v_s^2\to<v^2>/3$ \cite{1986R}. The correct expression for $<v^2>$ can be obtained using the method described above:
\begin{equation}
\label{vdisp}
<v^2>=
\left\{
\begin{array}{ll}
c^2 & z>z_{nr}
,\\
12\frac{\eta_R(5)}{\eta_R(3)}\left(\frac{kT_{\nu0}}{m_{\nu}}\right)^2
\frac{B(\xi)}{A(\xi)} & z<z_{nr}
,
\end{array}\right.
\end{equation}
where $T_{\nu0}=1.97\,$K is the present temperature of neutrinos.

\begin{figure}[ht]
\begin{center}
\includegraphics[width=\hsize,clip]{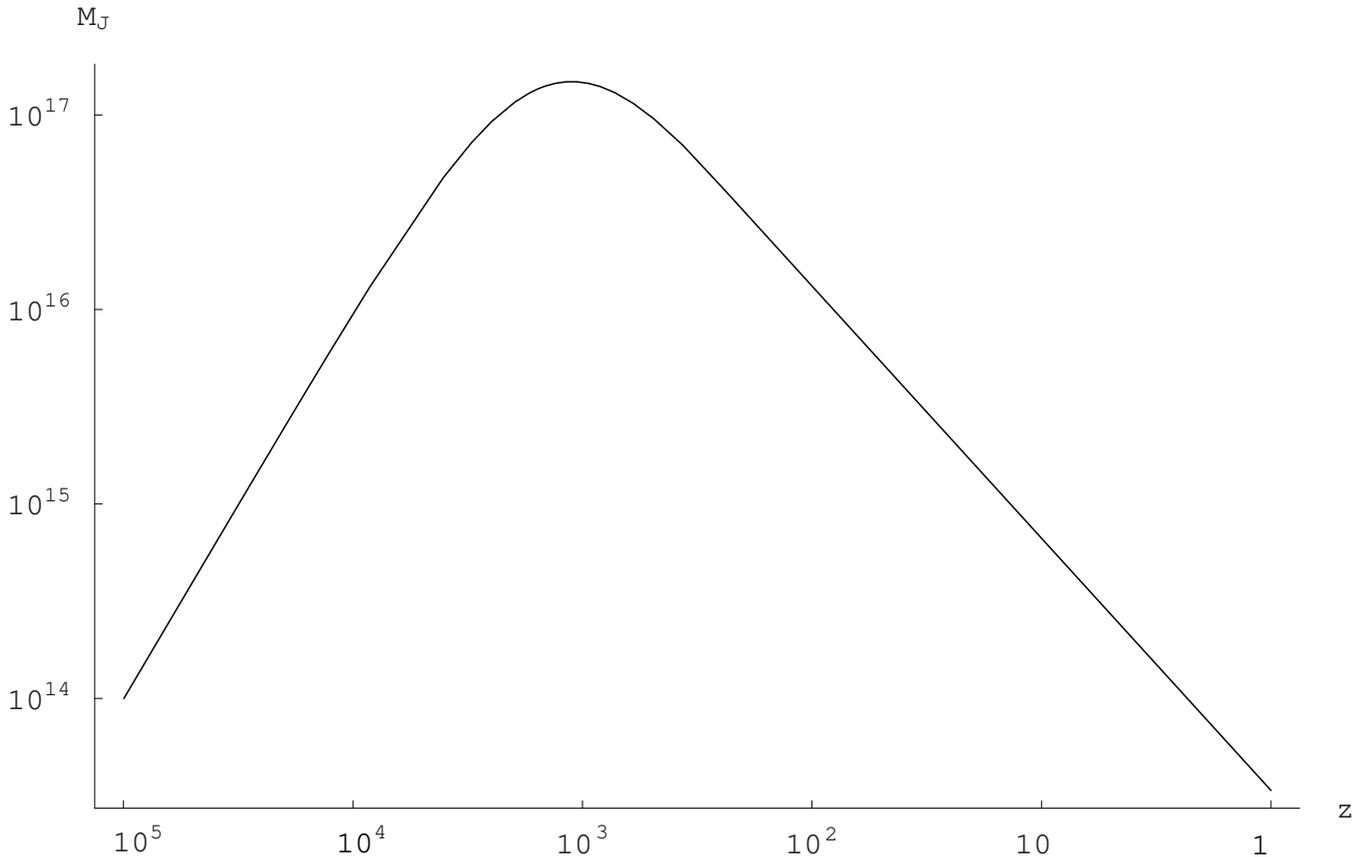}
\caption{The Jeans mass dependence on redshift for neutrinos with mass $m_{\nu}=2.5$eV and degeneracy parameter $\xi=2.5$.}
\label{fig:jeansz}
\end{center}
\end{figure}

As a result, the Jeans mass grows in the UR regime and decreases in the NR regime \cite{BES}. The evolution of the Jeans mass of neutrinos for $m_\nu=2.5\,$eV and $\xi=2.5$ with redshift $z$ is described by Fig.~\ref{fig:jeansz}.
It is clear that for such values of the neutrino mass the peak of the Jeans mass lies above $10^{17}\,M_\odot$ and the corresponding comoving Jeans length is $\lambda_0>100\,$Mpc. On the other hand, the value of the Jeans mass today is still larger than the mass of massive galaxy $10^{12}\,M_\odot$.

Finally, the maximum value of Jeans mass at the moment (\ref{rs}) is \cite{1986R}
\begin{equation}
\label{Jm}
M_J(z_{nr})=1.475 \;
10^{17}M_{\odot}g_{\nu}^{-\frac{1}{2}}N_{\nu}^{-\frac{1}{2}}
(\frac{m_{\nu}}{10eV})^{-2}A(\xi)^{-\frac{5}{4}}B(\xi)^{\frac{3}{4}}
.
\end{equation}

\begin{figure}[ht]
\begin{center}
\includegraphics[width=\hsize,clip]{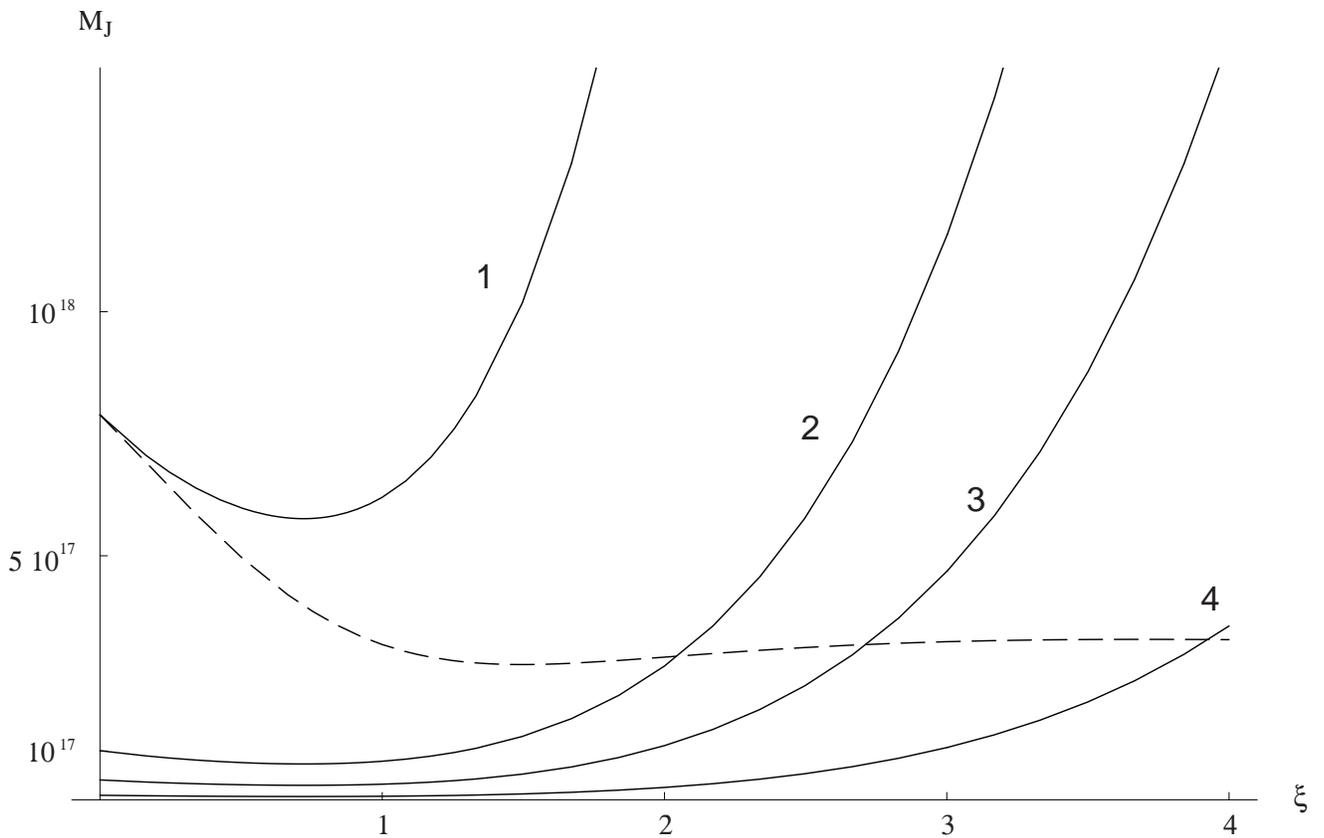}
\caption{The Jeans mass dependence on the degeneracy parameter with a fixed value of energy density, curves (1--4). Curve (1) corresponds to energy density $\Omega_{\nu}=0.11$. Curve (2) corresponds to $\Omega_{\nu}=0.3$.  Curve (3) represents the neutrino energy density $\Omega_{\nu}=0.5$ and finally curve (4) gives Jeans mass for $\Omega_{\nu}=1$. The dashed line represents Jeans mass dependence on the degeneracy parameter with fixed neutrino mass $m_{\nu}=2.5\,$eV.}
\label{fig:jeansm}
\end{center}
\end{figure}
The peak of the Jeans mass as a function of the degeneracy parameter for different fixed values of the energy density as well as with constant mass $m_{\nu}=2.5\,$eV is shown at Fig.\ref{fig:jeansm}.

By comparing different curves with a fixed value of $\xi$ one can find the well known result that the Jeans mass increases with decreasing of neutrino mass. With the growth of degeneracy parameter, however, the neutrino mass decreases in the beginning, and its different values correspond to different points at the same curve.

The space above the dashed line at Fig.~\ref{fig:jeansm} represents the region in which the neutrino mass is less than $2.5\,$eV. It iss interesting to note that this value of $m_{\nu}$ is still sufficient to get  $\Omega_{\nu}=1$ with $\xi\approx4$.

\subsection{Subsequent fragmentation model}

In this section we will describe a model of structure formation which explains  the observed fractal distribution matter. The key point of this model is the existence of upper and lower cutoffs in the fractal. The upper cutoff appears due to causality of the elementary cell, and the lower cutoff corresponds to the time when the dark matter ceases to dominate in the formation of the structures \cite{chaos}.

\subsubsection{Nonlinear model of spherical collapse}

Following Ruffini et al. \cite{1990R} we consider the first spherical perturbation in the Friedmann Universe. For simplicity we neglect the interaction between neighboring perturbations, and treat a given elementary cell formed at the epoch $z_{nr}$ as a spherically symmetric region. If the mass of an elementary cell is sufficiently larger than the Jeans mass, the mass density will dominate the pressure $p\ll\rho$ and then the sphere will expand freely from the expansion of the Friedmann background Universe. Then the equations governing the dynamical evolution of this shell can be obtained by solving the Einstein field equations in the Friedmann Universe in the metric
\begin{equation}
ds^2=-dt^2+e^{2\Lambda(\chi,t)}d\chi^2+r^2(\chi,t)(d\theta^2+sin^2\theta d\phi^2)
,
\end{equation}
where $\chi$ is a comoving radial coordinate and $r(\chi,t)$ is the radius of a 2-sphere. This metric is very similar to the Robertson-Walker one, because the spherical overdense (underdense) region behaves like a closed (open) sub-universe.

The solution of the Einstein equations with the above metric gives an equation for the total energy of a test particle:
\begin{equation}
\label{FirstInt}
\dot{r_1}^2-\frac{2Gm_1{\chi}}{r_1}+\epsilon_1(\chi)=0
,
\end{equation}
where $m_1(\chi)$ and $\epsilon_1(\chi)$ are respectively the gravitational mass and the negative total energy of the test particle at radius $r_1$ of a 2-sphere representing the first perturbed region.

We assume that $\epsilon_1(\chi)$ is constant in time since there is no dissipation and that at the moment of separation from the expansion of the background the expansion rates are the same $(\dot{a}/a)_1=(\dot{r}/r)_1$. If the mean mass density of the sphere is defined by
\begin{equation}
\bar{\rho_1}(t)\equiv\frac{m_1(\chi)}{\frac{4\pi}{3}r_1^3(\chi,t)}
,
\end{equation}
then we can write the energy $\epsilon_1(\chi)$ as
\begin{equation}
\epsilon_1(\chi)
=\frac{8\pi}{3}G\rho_U(t_1)R_1^2\bar\delta_1
=(H_1R_1)^2\bar\delta_1
.
\end{equation}
Here we have assumed a Friedmann Universe with $k=0$ as a background. $R_1=r(\chi, t_1),\,\bar\rho_1(t_1)$, and $\rho_U(t_1)$ are respectively the radius of the sphere, its mean density, and the mean density of the Universe at the epoch of separation $t=t_1$. Furthermore, $\bar\delta_1\equiv\bar\rho_1(t_1)/\rho_U(t_1)-1$ and $H_1$ is the Hubble parameter at $z=z_1$.
Eq.~(\ref{FirstInt}) can be solved in parametric form: $r_1=r_1(\Theta_1),\,t=t(\Theta_1)$, where the parameter $\Theta_1$ is a conformal time which has an initial value $\theta_1$ related to the initial value of the density contrast $\bar\delta_1$ by
\begin{equation}
\theta_1
=\frac{1}{\sqrt{\epsilon}} \arccos\left(\frac{1+\bar\delta_1
        -2\epsilon\bar\delta_1}{1+\bar\delta_1}\right)
,
\end{equation}
where $\epsilon=-1,\,+1$ corresponds to underdense and overdense regions respectively, being the analog of the curvature parameter $k$ in the usual Robertson-Walker metric.
Using the relation between cosmic time $t$ and redshift $z$ in a flat ($k=0$) Friedmann Universe, we arrive at the following expression for $z$ as a function of $\Theta$:
\begin{equation}
1+z=(1+z_1)\left[1+\frac{3}{4} \left(\frac{1+\bar\delta_1}{\bar\delta_1^{3/2}}\right)
(\Theta_1-\sin\Theta_1-\theta_1+\sin\theta_1)\right]^{-2/3}
.
\end{equation}
At the same time, an expression for $\bar\delta_1$ can be obtained:
\begin{equation}
\bar\delta_1(t)
=8\left(\frac{1+z_1}{1+z}\right)^3\frac{\bar\delta_1^3}{(1+\bar\delta_1)^2}
         (1-\cos\Theta_1)^{-3}-1
.
\end{equation}
This set of equations determines the evolution of the elementary cell.

\subsubsection{Successive fragmentation}

After the separation of the elementary cell from the Hubble flow, the Jeans mass will continue to fall and perturbations of smaller masses will be able to detach themselves from the expansion either of the cosmological background or of the larger parent elementary cell. The perturbed regions in the background will follow the evolution given above, but the perturbed regions in the parent elementary cell clearly follow a different evolution.

Here we discuss the simplified case of a spherically symmetric pertubation separating from the expansion flow of its parent cell. We can proceed exactly as we did for the initial elementary cell perturbations, remembering, however, that now the perturbation detaches itself from the still expanding parent cell and not from the Hubble flow. Thus the density and evolutionary state of the parent cell, and not the one of the Friedmann background, will enter into our calculation of the evolution of successive perturbations.

For the perturbation of $n$ we thus have
\begin{equation}
\bar\delta_n(t)
=8\left(\frac{1+z_n}{1+z}\right)^3\frac{(\bar\delta_n+\Delta_n)^3}
       {(1+\bar\delta_n)^2}(1-\cos\Theta_n)^{-3}-1
,
\end{equation}
where $\Delta_n$ is a parameter in which the whole history of the previous fragmentation is summarized. It has the expression
\begin{equation}
\Delta_n
=1-4\left(\frac{1+z_{n-1}}{1+z_n}\right)^3\frac{(\bar\delta_{n-1}
+\Delta_{n-1})^3}{(1+\bar\delta_{n-1})^2}
\frac{\sin^2\Theta^n_{n-1}}{(1-\cos\Theta^n_{n-1})^4}
,
\end{equation}
where $\Theta^n_{n-1}$ is the value of conformal time when the $n$-th fragmentation appears inside the ($n-1$)-th one, and $\Delta_1=0$ for the first fragmentation.

\subsubsection{The fractal model}

To explain in a simple way the mechanism of this model, we start with a simple twofold scenario in which each condensation gives birth to two daughter condensations as soon as the value of the Jeans mass of the parent cell drops to half of the initial value. In short, we take a condensation of mass $M_J$, then reach the redshift at which the Jeans mass has become $M_J/2$. At that epoch inside the initial condensation two new daughter condensations originate, each of mass $M_J/2$. As we continue this process, we will have four granddaughter condensations, then eight, sixteen and so on. This process can easily be generalized to the occurence of $N$ fragments at each step.

Since the mass of the initial condensation as well as the behaviour of the Jeans mass and length are given, the only free parameter at each successive step is the amplitude of the perturbation $\bar\delta_n$. Our goal is to reproduce at the end of the fragmentation process the expected fractal distribution and, therefore, to select at each step the suitable perturbation $\bar\delta_n$ for this purpose.
It has been shown \cite{Mandelbrot} that such a process leads to a system with fractal dimension $D_F$ given by
\begin{equation}
\label{fracdim}
D_F=\frac{\log N}{\log \lambda}
,
\end{equation}
where $\lambda=r_{i-1}/r_i$ is a constant. In our case we assume $D_F=1.2$ \cite{1988CR}.

In the spherical model decribed above the density need not be uniformly distributed inside the radius $r_n$: any spherically symmetric perturbation will clearly evolve at a given radius $r_n$ in the same way as a uniform sphere containing the same mass. We assume then without loss of generality that
\begin{equation}
\delta_n(r)
=
\left\{
\begin{array}{ll}
\bar\delta_n, & r<r_n ,\\
0, & r>r_n
.
\end{array}
\right.
\end{equation}

\begin{figure}
\begin{center}
\includegraphics[width=\hsize,clip]{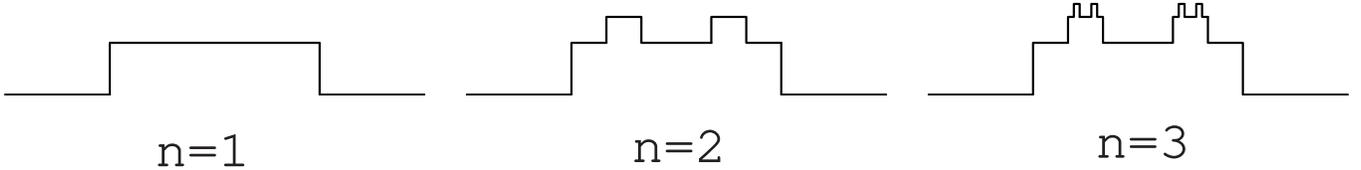}
\caption{Qualitative illustration of subsequent fragmentation process in the spherical model. The final picture obtained from such a mechanism looks like a typical fractal.}
\label{fig:frac}
\end{center}
\end{figure}

The process of fragmentation is shown qualitatively in Fig.~\ref{fig:frac}. This picture is reminiscent of a fractal. Note that the position of each new fragment inside previous one is not important. Moreover, predictions of the model are quite insensitive to the number $N$. The key point is that even for a random number of fragments appearing at each step the resulting density distribution is still a fractal. One difficulty with this model is that it predicts too large a density contrast $\bar\delta_0$ today at the galactic scale.

In order to avoid such high values of the density contrast the authors introduced a suitable lagging time factor $\tau$. This factor is a function of $N$ that introduces a time delay in the formation of each daughter condensation. It was shown that a valid phenomenological relation leading to realistic values of density contrast today is
\begin{equation}
\tau(N)=\frac{N^2}{4}+\frac{3}{4}N
.
\end{equation}
Clearly, the existence of the lagging factor $\tau$ is not in contradiction with the Jeans instability picture and only means that the fragmentation occurs somewhat later than the time at which the necessary condition is fulfilled.

In the model under consideration the only free parameters are the initial density contrasts $\bar\delta_n$ at every step. These are chosen in such a way that Eq.~(\ref{fracdim}) is satisfied at each step. Thus the natural question arises, what is the form of the initial spectrum? It can be obtained by following backward in time the evolution of perturbations.
The result is quite surprising and simple. The spectral index of initial spectrum  at $z_{nr}$ (see (\ref{insp})) is $n=0$, which corresponds to white noise.
Note also that agreement with the observed CBR anisotropy can be obtained within the framework of this model \cite{1992FR}.


\end{document}